%% file: root.tex
\documentclass{article}

\usepackage[utf8]{inputenc}
\usepackage{graphicx}
\usepackage{amsmath}
\usepackage{amsfonts}
\usepackage{amsthm}
\usepackage{bm}
\usepackage[font=small,skip=0pt]{caption}
\usepackage{subcaption}
\usepackage{psfrag}
\usepackage{xcolor} 

\newtheorem{theorem}{Theorem}
\newtheorem{lemma}{Lemma}

\newdimen\figrasterwd
\figrasterwd\textwidth

\title{Distributed Kalman Filters for Relative Formation Control of Multi-agent systems}

\author{Martijn van der Marel and Raj Thilak Rajan\thanks{The authors are with the Faculty of EEMCS, Delft University of Technology (TUD), The Netherlands. This work is partially funded by the European Leadership Joint Undertaking (ECSEL JU), under grant agreement No 876019, the ADACORSA project - ”Airborne Data Collection on Resilient System Architectures.”}\thanks{Corresponding author: R.T.Rajan, R.T.Rajan@tudelft.nl}}

\begin{document}

\maketitle

\begin{abstract}
    Formation control (FC) of multi-agent plays a critical role in a wide variety of fields. In the absence of absolute positioning, agents in FC systems rely on relative position measurements with respect to their neighbors. In distributed filter design literature, relative observation models are comparatively unexplored, and in FC literature, uncertainty models are rarely considered. In this article, we aim to bridge the gap between these domains, by exploring distributed filters tailored for relative FC of swarms. We propose statistically robust data models for tracking relative positions of agents in a FC network, and subsequently propose optimal Kalman filters for both centralized and distributed scenarios. Our simulations highlight the benefits of these estimators, and we identify future research directions based on our proposed framework.
\end{abstract}
\input{sections/1_introduction}

\input{sections/2_preliminaries}
\input{sections/3_data_model}

\input{sections/4_simulations}

\input{sections/5_summary}

\newpage
\bibliographystyle{IEEEbib}
\bibliography{ref}


\end{document}

%% file: sections/1_introduction.tex
\section{Introduction}\label{sec:introduction} 
Formation control (FC) of multi-agent systems plays a crucial role in various applications, for e.g., in satellite interferometry \cite{Liu2018}, UAV formation flight \cite{Zou2018}, and underwater sensing networks \cite{Bechlioulis2019}. Traditionally, the task of keeping a multi-agent system in the desired formation is performed centrally, however, the inherent distributed nature of multi-agent systems naturally invites a decentralized architecture. One class of distributed methods focus on solving a global optimization problem by distributing the problem over the agents \cite{Raffard2004}, and other solutions are based on behavior-based algorithms \cite{Balch1998} or leader-follower architectures \cite{Chen2021}. Motivated by the need for FC in GNSS-denied environments (e.g., indoors, in space or underwater), formations in the context of relative navigation have been investigated, where the agents navigate based on their relative dynamics with respect to other mobile agents or objects \cite{Cheng2015}. One approach to tackle this challenge is through the extension of graph Laplacian-based consensus algorithms \cite{Han2015}. For example, in \cite{Lin2016} a multi-agent system with single-integrator dynamics is considered, and closed form solution for a local control law is proposed, which guarantees the convergence to an affine formation or a rigid formation \cite{Lin2016, Park2018}.
\par
A relatively unexplored area within the domain of relative FC is that of statistical uncertainty. In practise, the dynamics of mobile agents and the measurements they make are corrupted by noise, which can be cast as a linear state space model, which motivates the need for designing distributed Kalman filters (KFs). A plethora of existing methods for distributed KFs focus on estimating a common environment state \cite{Olfati-Saber2005,Talebi2019}, but in case of FC, the agents must track their own dynamical states, which can be seen as subsets of the global state variables. More recently, distributed KFs have been proposed for linear FC systems \cite{Marelli2018} and \cite{Viegas2018}, however, these algorithms require absolute state measurements,  which we aim to overcome in this work. 

\par
\textit{Notation:} Vectors and matrices are represented by lowercase and uppercase boldface letters respectively. $A_{ij}$ represents the element on the $i$th row and $j$th column of the matrix $\mathbf{A}$. Sets and graphs are represented using calligraphic letters e.g., $\mathcal{A}$.  A vector of length $N$ of all ones and zeros are denoted by $\mathbf{1}_N$ and $\mathbf{0}_N$ respectively. An identify matrix of size $N$ is denoted by $\mathbf{I}_N$. The Kronecker product is $\otimes$,  $\mathrm{tr}(\cdot)$ is the trace, $\mathbb{E}(\cdot)$ is the expectation, and $\mathrm{bdiag}(\mathbf{A}_i)_{i\in\mathcal{S}}$ is a block diagonal matrix with blocks $\mathbf{A}_i$ $\forall i\in\mathcal{S}$.

%% file: sections/2_preliminaries.tex
\subsection{Related work}\label{sec:preliminaries} 

Consider a swarm comprising of $N$ homogeneous mobile agents moving in $D$-dimensional space. The sensing capabilities of the agents are described by the bidirectional sensing graph $\mathcal{G} = (\mathcal{V},\mathcal{E})$, where the nodes of the graph $\mathcal{V}=\{1,\dots,N\}$ denote the agents. The edges represent the sensing links: $(i,j)\in\mathcal{E}$ which  implies the agent $i$ can measure its relative position with respect to agent $j$. The neighborhood of a node $i$ is denoted by $\mathcal{N}_i$, i.e. $j\in\mathcal{N}_i$ implies $(i,j)\in\mathcal{E}$. The position of agent $i$ in a $D$-dimensional Euclidean space is denoted by the vector $\mathbf{z}_i\in\mathbb{R}^D$. 
We consider the mobile agents to be governed by single-integrator dynamics i.e., \begin{eqnarray}\label{eq:local_dynamics_nonoise}
    \dot{\mathbf{z}}_i &=& \mathbf{u}_i \\ 
    \label{eq:local_control_law}
    \mathbf{u}_i &=& \displaystyle\sum_{j\in\mathcal{N}_i}l_{ij}(\mathbf{z}_i-\mathbf{z}_j)
\end{eqnarray} where $\mathbf{u}_i$ represents the control input of agent $i$. Here, the control input for individual agents is a weighted sum of relative positions, where the weights $l_{ij}$ can be considered as elements of the generalized graph Laplacian. These weights depend on the desired formation configuration, and a convex optimization program for designing the weights is additionally provided in \cite{Lin2016}. Given a system of mobile agents, governed by the dynamics (\ref{eq:local_dynamics_nonoise}) and following the control law (\ref{eq:local_control_law}), the convergence to any desired affine formation is possible, if and only if the sensing graph $\mathcal{G}$ is universally rigid \cite{Lin2016}.  In various applications, a rigid formation is preferable to an affine formation as it preserves inter-agent distances, which can be achieved by the appointment of $D+1$ leader agents in the swarm who may follow e.g. the local control law proposed in \cite{Park2018} to achieve this goal. It is worth noting that measurement uncertainties were not modelled in these approaches.
\par
In this paper, we aim to statistically model the uncertainty in the swarm, to improve the robustness of existing FC solutions. In real-world FC applications, there are a various sources of uncertainty which affect the dynamical system. These can be broadly categorized into two groups, i.e., observation noise originating from the sensors, and uncertainties originating from the environment e.g., wind for UAV formations.




%% file: sections/3_data_model.tex
\section{Local State-Space Model} \label{sec:data_model} 

We begin by considering a single edge of the sensing graph, and based on the proposed data models, we develop two local estimators for estimating the relative position of the agents, which drive the control law (\ref{eq:local_control_law}). 

\subsection{Maximum likelihood estimation (MLE)}
Consider a single edge $(i,j)$ of the sensing graph, where the relative position (or \textit{edge state}) of the agent $i$ with respect to a neighbour $j$ at time index $k$ is given by $\mathbf{z}^{ij}_k = \mathbf{z}^i_k - \mathbf{z}^j_k$. Let agent $i$ be able to measure the relative position of its neighbor $j$, which can be achieved using numerous methods \cite{buehrer2018collaborative,rajan2019relative}. Consider that $T$ i.i.d. measurements are made at time index $k$, which are corrupted with additive noise, then the relative position measurements are given by,\begin{equation}\label{eq:observation_model}
    \mathbf{y}_k^{ij}=\mathbf{H}\mathbf{z}_k^{ij}+\mathbf{v}_k^{ij} 
\end{equation} where we assume linear measurement model with $\mathbf{y}^{ij}_k$ as the measurement vector and  $\mathbf{H}=\mathbf{1}_T\otimes\mathbf{I}_D$ is the observation matrix. We consider the noise plaguing the measurements to be normally distributed i.e.,$\mathbf{v}_k^{ij}\sim\bm{\mathcal{N}}(\mathbf{0}_{TD},\mathbf{I}_T\otimes\mathbf{R}_{ij})$,where the noise is correlated with regard to dimensions but uncorrelated in time. Given the measurements $\mathbf{y}^{ij}_k$, the MLE for the relative position vector at time-index $k$ can be obtained by maximizing the probability density function of the measurement vector. i.e., $\hat{\mathbf{z}}_k^{ij} = (\mathbf{H}^\top\Bar{\mathbf{R}}^{-1}_{ij}\mathbf{H})^{-1}\mathbf{H}^\top\Bar{\mathbf{R}}^{-1}_{ij}\mathbf{y}_k^{ij}$ $= T^{-1}\mathbf{H}^\top\mathbf{y}_k^{ij}$ where we introduced $\Bar{\mathbf{R}}_{ij}= \mathbf{I}_T\otimes\mathbf{R}_{ij}$.


\begin{figure}
    \centering
    \includegraphics[width=0.8\linewidth]{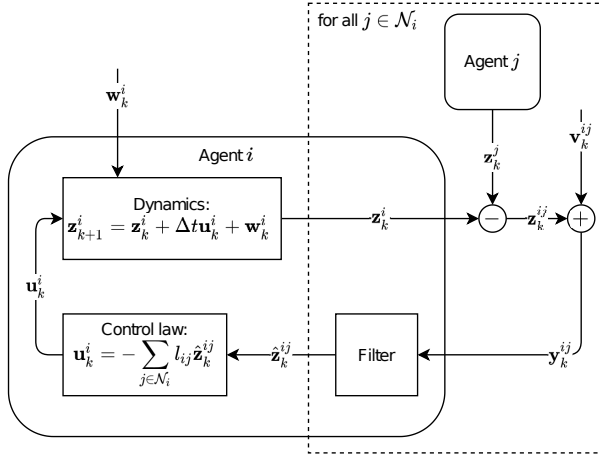}
    \caption{Proposed local data model from the perspective of agent $i$.}
    \label{fig:block}
\end{figure}

\subsection{Edge-based Relative Kalman filter (RKF)} The estimation accuracy of the MLE can be further improved over time, and subsequently the discrete-time dynamics of the edge state from the agent dynamics (\ref{eq:local_dynamics_nonoise}) can be deduced as \begin{equation}\label{eq:dynamics_model}
    \mathbf{z}_{k+1}^{ij} = \mathbf{z}_k^{ij} + \Delta t(\mathbf{u}_k^i-\mathbf{u}_k^j) + \mathbf{w}_k^{ij},\quad \mathbf{w}_k^{ij}\sim\bm{\mathcal{N}}(\mathbf{0}_{D},\mathbf{Q}_{ij})
\end{equation} where $\Delta t$ denotes the time step of the filter, where we assume the process noise $\mathbf{w}_k^{ij}$ to be Gaussian with a known covariance. The block diagram in Figure \ref{fig:block} visualizes the data model with the noisy measurements and dynamics. Since both (\ref{eq:observation_model}) and (\ref{eq:dynamics_model}) are Gauss-Markov models, the KF is optimal for estimating the time-varying edge state or relative position. For the rest of this paper, we assume the initial positions $\mathbf{z}_0^i$ of all agents $i=1,\dots,N$ to be drawn from the  arbitrary distribution $\bm{\mathcal{N}}(\bm{\mu},\mathbf{P})$. Subsequently, the initialization, prediction and update steps of the RKF are as follows. \begin{subequations}
\label{eq:EKF}
\begin{align}
        \mathbb{E}(\mathbf{z}_0^{ij}) &=
        \mathbb{E}(\mathbf{z}_0^{i})-\mathbb{E}(\mathbf{z}_0^j)  = \mathbf{0}_D, \\
        \bm{\Sigma}_{0|0}^{ij} &= 
        \mathbb{E}\left((\mathbf{z}_0^{i}-\mathbf{z}_0^{j})(\mathbf{z}_0^{i}-\mathbf{z}_0^{j})^\top\right)=
        2\mathbf{P} \\
        \mathbf{\hat{z}}_{k+1|k}^{ij} &= 
            \mathbf{\hat{z}}_{k|k}^{ij} + \Delta t(\mathbf{u}_k^i-\mathbf{u}_k^j)\\
        \bm{\Sigma}_{k+1|k}^{ij} &= 
            \bm{\Sigma}_{k|k}^{ij}+\mathbf{Q}_{ij}\label{eq:ekf_pred_cov} \\
        \mathbf{K}_k^{ij} &= 
        \bm{\Sigma}_{k|k-1}^{ij}\mathbf{H}^\top\big(\mathbf{H}\bm{\Sigma}_{k|k-1}^{ij}\mathbf{H}^\top+\mathbf{I}_T\otimes\mathbf{R}_{ij}\big)^{-1} \label{eq:ekf_update_gain} \\
        \mathbf{\hat{z}}_{k|k}^{ij} &= \mathbf{\hat{z}}_{k|k-1}^{ij} + \mathbf{K}_k^{ij}\big(\mathbf{y}_k^{ij}-\mathbf{H}\mathbf{\hat{z}}_{k|k-1}^{ij}\big) \label{eq:ekf_update_mean} \\
        \bm{\Sigma}_{k|k}^{ij} &= \big(\mathbf{I}_D-\mathbf{K}_k^{ij}\mathbf{H}\big)\bm{\Sigma}_{k|k-1}^{ij} \label{eq:ekf_update_cov}
    \end{align}
\end{subequations} 

\section{Global State-Space Model} \label{sec:data_model_global} We now extend the edge-based solutions to a global state-space model, and subsequently propose optimum centralized and distributed KFs.
 
\subsection{Centralized Relative Kalman filter (CRKF)}
To design a solution for the entire network, we introduce the vectors $\mathbf{z}_k=\{\mathbf{z}_k^i\}_{i=1}^N\in\mathbb{R}^{ND}$ and $\mathbf{u}_k = \{\mathbf{u}_k^i\}_{i=1}^N\in\mathbb{R}^{ND}$, which concatenate the absolute positions and the control inputs of all agents respectively. The aggregated agent dynamics can then be written as 
\begin{equation}\label{eq:global_agent_dynamics}
    \mathbf{z}_{k+1}=\mathbf{z}_k+\Delta t\mathbf{u}_k+\mathbf{w}_k,\quad\mathbf{w}_k\sim\bm{\mathcal{N}}(\mathbf{0}_{ND},\mathbf{Q})
\end{equation}
Here, the vector $\mathbf{w}_k=\{\mathbf{w}_k^i\}_{i=1}^N$ represents the spatial disturbances acting on the agent dynamics, and the covariance matrix $\mathbf{Q}$ is therefore considered to be a full matrix.  Observe that (\ref{eq:global_agent_dynamics}) denotes the agent dynamics containing the absolute positions, while the control input is driven by the relative positions w.r.t the neighbouring agents (\ref{eq:local_control_law}). Thus, to map the absolute position vector $\mathbf{z}_k$ to the relative positions, we exploit the known incidence matrix $\mathbf{B} = \{B_{il}\}$ of the Graph, whose entries are $+1$ when $l$ is directed towards $i$, $-1$ when $l$ is directed away from $i$, and $0$ otherwise. Following the convention in \cite{Viegas2018}, the columns of the incidence matrix $\mathbf{B}$ are grouped per node, starting with all edges directed towards node $1$ and ending with edges directed towards node $N$. Let $\mathbf{x}_k = \left(\mathbf{B}^\top\otimes\mathbf{I}_D\right)\mathbf{z}_k = \Bar{\mathbf{B}}^\top\mathbf{z}_k$ be an edge state vector, the left multiplying (\ref{eq:global_agent_dynamics}) by $\Bar{\mathbf{B}}^\top$, the discrete-time edge dynamics for the entire network is \begin{equation}\label{eq:global_edge_dynamics}
    \mathbf{x}_{k+1} = \mathbf{x}_k + \Delta t\Bar{\mathbf{B}}^\top\mathbf{u}_k + \bar{\mathbf{w}}_k
\end{equation} where $\bar{\mathbf{w}}_k = \bar{\mathbf{B}}^\top\mathbf{w}_k \sim\bm{\mathcal{N}}(\mathbf{0}_{ND},\Bar{\mathbf{B}}^\top\mathbf{Q}\Bar{\mathbf{B}})$. Note that for the single-edge dynamical model in (\ref{eq:dynamics_model}), the covariance matrix can be retroactively defined as $\mathbf{Q}_{ij}= \Bar{\mathbf{B}}_{ij}^\top\mathbf{Q}\Bar{\mathbf{B}}_{ij}$ where $\Bar{\mathbf{B}}_{ij}=\mathbf{b}_{ij}\otimes\mathbf{I}_D$ where $\mathbf{b}_{ij}$ is the column of the incidence matrix representing the edge $(i,j)$. Furthermore, the observation model (\ref{eq:observation_model}) for the global system is 
\begin{equation}\label{eq:global_observation_model}
    \mathbf{y}_k = \Bar{\mathbf{H}}\mathbf{x}_k + \mathbf{v}_k,\quad\mathbf{v}_k\sim\bm{\mathcal{N}}(\mathbf{0}_{MTD},\mathbf{R})
\end{equation}
with $\Bar{\mathbf{H}}=(\mathbf{I}_M\otimes \mathbf{H})$ where $M$ is the total number of edges in the network, and $\mathbf{R}=\mathrm{bdiag}(\mathbf{I}_T\otimes\mathbf{R}_{ij})_{(i,j)\in\mathcal{E}}$ a block diagonal matrix with the ordering of the edges equivalent to that of the incidence matrix. Given  (\ref{eq:global_edge_dynamics}) and (\ref{eq:global_observation_model}), we propose the CRKF for the entire network as follows. \begin{subequations} \label{eq:CRKF}
\begin{align}
        \label{eq:cekf_init_mean}
        \hat{\mathbf{x}}_{0|0} &= 
        \Bar{\mathbf{B}}^\top(\mathbf{1}_N\otimes\bm{\mu})=
        \mathbf{0}_{2MD}\\
        \label{eq:cekf_init_cov}
        \bm{\Sigma}_{0|0} &= 
        \Bar{\mathbf{B}}^\top(\mathbf{I}_N\otimes\mathbf{P})\Bar{\mathbf{B}} = \mathbf{B}^\top\mathbf{B}\otimes\mathbf{P}\\
        \label{eq:cekf_predict_mean}
        \hat{\mathbf{x}}_{k+1|k} &= 
        \hat{\mathbf{x}}_{k|k} + \Delta t\Bar{\mathbf{B}}^\top\mathbf{u}_k\\
        \label{eq:cekf_predict_cov}
        \bm{\Sigma}_{k+1|k} &= 
        \bm{\Sigma}_{k|k} + \Bar{\mathbf{B}}^\top\mathbf{Q}\Bar{\mathbf{B}}\\
        \label{eq:cekf_update_gain}
        \mathbf{K}_k &= 
        \bm{\Sigma}_{k|k-1}\Bar{\mathbf{H}}^\top(\Bar{\mathbf{H}}\bm{\Sigma}_{k|k-1}\Bar{\mathbf{H}}^\top+\mathbf{R})^{-1}\\
        \label{eq:cekf_update_mean}
        \hat{\mathbf{x}}_{k|k} &= 
        \hat{\mathbf{x}}_{k|k-1} + \mathbf{K}_k(\mathbf{y}_k-\Bar{\mathbf{H}}\hat{\mathbf{x}}_{k|k-1})\\
        \label{eq:cekf_update_cov}
        \bm{\Sigma}_{k|k} &= 
        (\mathbf{I}_{MD}-\mathbf{K}_k\Bar{\mathbf{H}})\bm{\Sigma}_{k|k-1}
\end{align}
\end{subequations} 




\subsection{Joint Relative Kalman filter (JRKF)} To alleviate the centralized collection of information and processing, we propose a distributed algorithm for the CRKF. Observe that the steps in (\ref{eq:CRKF}) are node separable when the Kalman gain matrix in (\ref{eq:cekf_update_gain}) is block diagonal i.e., it can be expressed as $\mathbf{K}_k = \mathrm{bdiag}(\mathbf{K}_k^i)_{i\in\mathcal{V}}$, where $\mathbf{K}_k^i\in\mathbb{R}^{M_iD\times M_iTD}$, where $M_i = |\mathcal{N}_i|$ denotes the cardinality of the neighbourhood of agent $i$. However, the key challenge is the distributed computation of the optimal block diagonal Kalman gain matrix. \footnote{More generally, for any sparsity structure constraint on the Kalman gain matrix, a solution is provided in \cite{Viegas2018}, however this requires a centralized computation.} To this end, we propose the following theorem, supported by a lemma.

\begin{theorem} A solution to the Kalman gain $\mathbf{K}_k^i\in\mathbb{R}^{M_iD\times M_iTD}$ can be computed locally by agent $i$ by solving the cost function \begin{equation}\label{eq:trace_1}
        \min_{\mathbf{K}_k^i,i\in\mathcal{V}}
        \quad \mathrm{tr}\left(\bm{\Sigma}_{k|k}\right)
        \quad \text{s.t.} \quad\mathbf{K}_k = \mathrm{bdiag}(\mathbf{K}_k^i)_{i\in\mathcal{V}}
\end{equation} which admits the following solution 
\begin{equation}\label{eq:opt_local_kg}
    \mathbf{K}_k^i = \bm{\Sigma}_{k|k-1}^i\mathbf{H}_i^\top\left(\mathbf{H}_i\bm{\Sigma}_{k|k-1}^i\mathbf{H}_i^\top+\mathbf{R}_i\right)^{-1}
\end{equation} where $\mathbf{H}_i = \mathbf{I}_{M_i} \otimes \mathbf{H}$
and $\mathbf{R}_i = \mathrm{bdiag}(\mathbf{R}_{ij})_{j \in \mathcal{N}_i}$.
\end{theorem} 

\begin{lemma}\label{le:tr}
Let $\mathbf{A} =\mathrm{bdiag}(\mathbf{A}_i)_{i=1}^N$, and consider $\mathbf{F}$ with the dimensions of $\mathbf{A}$, then $\mathrm{tr}(\mathbf{AFA^\top})$ $=$
$\sum_{i=1}^{N} \mathrm{tr}(\mathbf{A}_i\mathbf{F}_{ii}\mathbf{A}_i^\top)$
where $\{\mathbf{F}_{11}, \mathbf{F}_{22}, \hdots, \mathbf{F}_{NN} \}$ are the block diagonal matrices of $\mathbf{F}$ \cite{Golub1996}.
\end{lemma} 

\begin{proof} Since minimizing the mean square error in the Kalman updates (\ref{eq:cekf_update_gain}) is equivalent to minimizing the trace of the posterior covariance, we have (\ref{eq:trace_1}). Now substituting the Joseph form of the posterior covariance, and introducing $\mathbf{\Phi} = \mathbf{I}_{MD}-\mathbf{K}_k\Bar{\mathbf{H}}$, we have
\begin{equation} \begin{split} \label{eq:trace_2}
        \min_{\mathbf{K}_k^i,i\in\mathcal{V}} 
        \mathrm{tr}\left( \mathbf{K}_k\mathbf{R}\mathbf{K}_k^\top\right) +  
        \mathrm{tr}\left( \mathbf{\Phi}\bm{\Sigma}_{k|k-1}\mathbf{\Phi}^\top\right) \\
        \quad \text{s.t.} \quad \mathbf{K}_k = \mathrm{bdiag}(\mathbf{K}_k^i)_{i\in\mathcal{V}}
        \end{split}
\end{equation} where the constraint guarantees that the Kalman gain matrix $\mathbf{K}_k$ is block diagonal. Similarly,  $\mathbf{\Phi}$ is block diagonal since $\Bar{\mathbf{H}}$ is block diagonal by definition. Now, by introducing $\mathbf{\Phi}_i = \mathbf{I}_{M_iD}-\mathbf{K}_k^i\mathbf{H}_i$ and by applying the Lemma on the second summand of (\ref{eq:trace_2}), we decouple the optimization problem as \begin{equation}\label{eq:trace_3}
    \min_{\mathbf{K}_k^i,i\in\mathcal{V}}  \mathrm{tr}(\mathbf{K}_k^i\mathbf{R}_i\mathbf{K}_k^{i\top}) +
    \mathrm{tr}(\mathbf{\Phi}_i \bm{\Sigma}_{k|k-1}^i\mathbf{\Phi}_i^\top) 
\end{equation} Observe that this optimization problem is only minimizing the trace of the local posterior covariance, similar to an unconstrained Kalman gain optimization problem, which has a solution given by (\ref{eq:opt_local_kg}).
\end{proof} In summary, we propose the JRKF as follows 
\begin{subequations}
    \begin{align}
        \label{eq:jekf_init}
            \hat{\mathbf{x}}_{0|0}^i &= \mathbf{0}_{M_iD}, \\
            \bm{\Sigma}_{0|0}^i &= \mathbf{B}_i^\top\mathbf{B}_i\otimes\mathbf{P} \\
            \label{eq:jekf_predict_mean}
            \hat{\mathbf{x}}_{k+1|k}^i &= 
            \hat{\mathbf{x}}_{k|k}^i + \Delta t\Bar{\mathbf{B}}_i^\top\mathbf{u}_k, \\
            \bm{\Sigma}_{k+1|k}^i &= 
            \bm{\Sigma}_{k|k}^i + \Bar{\mathbf{B}}_i^\top\mathbf{Q}\Bar{\mathbf{B}}_i \\
            \label{eq:jekf_update_gain}
            \mathbf{K}_k^i &= \bm{\Sigma}_{k|k-1}^i\mathbf{H}_i^\top\left(\mathbf{H}_i\bm{\Sigma}_{k|k-1}^i\mathbf{H}_i^\top+\mathbf{R}_i\right)^{-1}\\
            \label{eq:jekf_update_mean}
            \hat{\mathbf{x}}_{k|k}^i &= \hat{\mathbf{x}}_{k|k-1}^i + \mathbf{K}_k^i\left(\mathbf{y}_k^i-\mathbf{H}_i\hat{\mathbf{x}}_{k|k-1}^i\right)\\
            \label{eq:jekf_update_cov}
            \bm{\Sigma}_{k|k}^i &= \left(\mathbf{I}_{M_iD}-\mathbf{K}_k^i\mathbf{H}_i\right)\bm{\Sigma}_{k|k-1}^i
    \end{align}
\end{subequations}
where $\Bar{\mathbf{B}}_i=\mathbf{B}_i\otimes\mathbf{I}_D$ and $\mathbf{B}_i$ is a submatrix of the incidence matrix $\mathbf{B}$ containing only the columns representing incoming edges of agent $i$. Unlike the RKF, which operates only over a single pairwise relative position, JRKF, jointly estimates the relative positions of multiple neighbours. Finally, note that for both the proposed distributed filters, the prediction steps of agents require the control inputs of neighboring agents, and hence local communication between agents is required.

%% file: sections/4_simulations.tex
\begin{figure}[t]
    \centering
    \begin{subfigure}[b]{0.45\linewidth}
         \centering
         \includegraphics[width=0.8\linewidth]{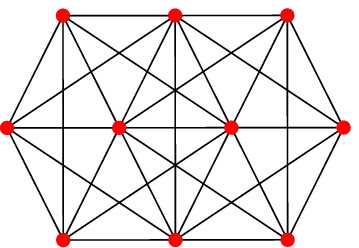}
         \caption{}
         \label{fig:framework}
     \end{subfigure}
     \begin{subfigure}[b]{0.45\linewidth}
         \centering
         \includegraphics[width=0.9\linewidth]{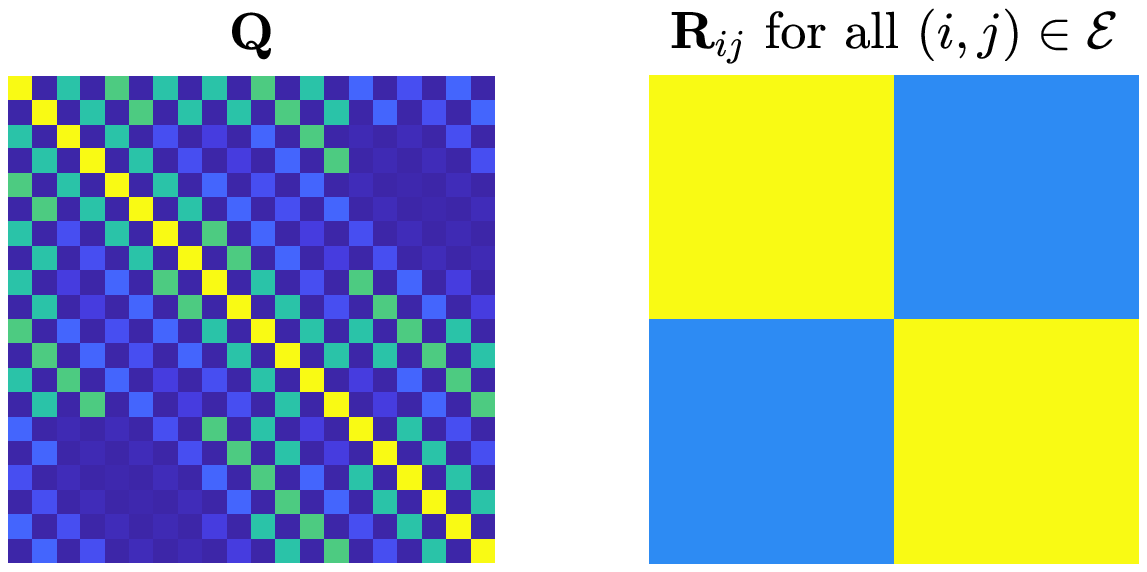}
         \caption{}
         \label{fig:covariances}
     \end{subfigure}
    \caption{(a) A formation of $N=10$ agents with $M=60$ bidirectional sensing links in $D=2$ dimensions. An illustration of the covariance matrices $\mathbf{Q}$ and $\mathbf{R_{ij}}$ used in the simulations.}
    \label{fig:sim_setup}
\end{figure}

\section{Simulations}\label{sec:simulations} We run simulations to quantify the performance of the proposed filters. We consider a set of mobile agents with single-integrator dynamics (\ref{eq:local_dynamics_nonoise}), following the control law (\ref{eq:local_control_law}). A subset of 3 nodes with an underlying complete subgraph follow the control law in \cite{Dimarogonas2009} instead, which preserves the inter-agent distances between these leader agents. The formation framework is visualized in Figure \ref{fig:framework}. The underlying sensing graph is universally rigid,  which guarantees convergence to a rigid formation in the noiseless, continuous-time case as per \cite{Lin2016}.Discrete-time simulations should approach this behavior with small time intervals $\Delta t=1$ milliseconds (ms). We use the uncertainty model in (\ref{eq:global_agent_dynamics}),  and the observation model (\ref{eq:global_observation_model}) with $T=10$, where the initial positions of agents are drawn i.i.d. from $\bm{\mathcal{N}}(\bm{\mu},\mathbf{P})$ with $\bm{\mu}=\mathbf{0}_D$ and $\mathbf{P}=\mathbf{I}_D$. We assume $\mathbf{Q}$ is constant over time, with diagonal elements $\sigma_w^2$ and off-diagonal elements in $[0,\sigma_w^2]$, and similarly $\mathbf{R}_{ij}$ are assumed equivalent for all edges $(i,j)$ with diagonal elements $\sigma_v^2$ and off-diagonal elements in $[0,\sigma_v^2]$. See \figurename\ \ref{fig:covariances}.

\begin{figure}[t]
    \centering
    \psfrag{Lin2016}[b]{\footnotesize Lin \textit{et al.} \cite{Lin2016}}
    \psfrag{MLE}[b]{\footnotesize MLE (proposed)}
    \psfrag{RKF}[b]{\footnotesize RKF (proposed)}
    \psfrag{JRKF}[b]{\footnotesize JRKF (proposed)}
    \includegraphics[width=0.6\linewidth]{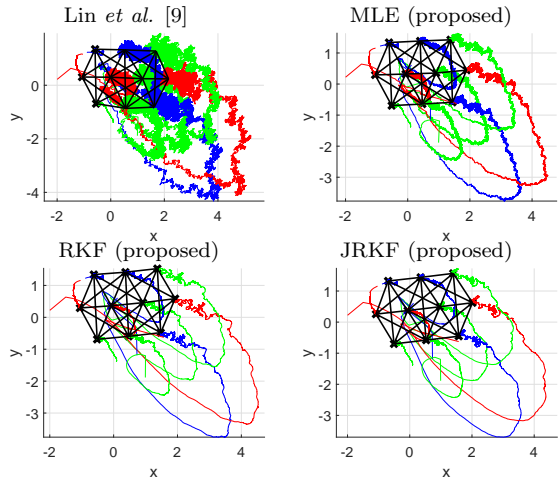}
    \caption{Paths of agents in $\mathbb{R}^2$ over time for the proposed estimators }
    \label{fig:paths}
\end{figure}

\begin{figure}[t]
    \begin{subfigure}[b]{0.49\linewidth}
        \centering
        \includegraphics[height=0.83\linewidth, width=1.1\linewidth]{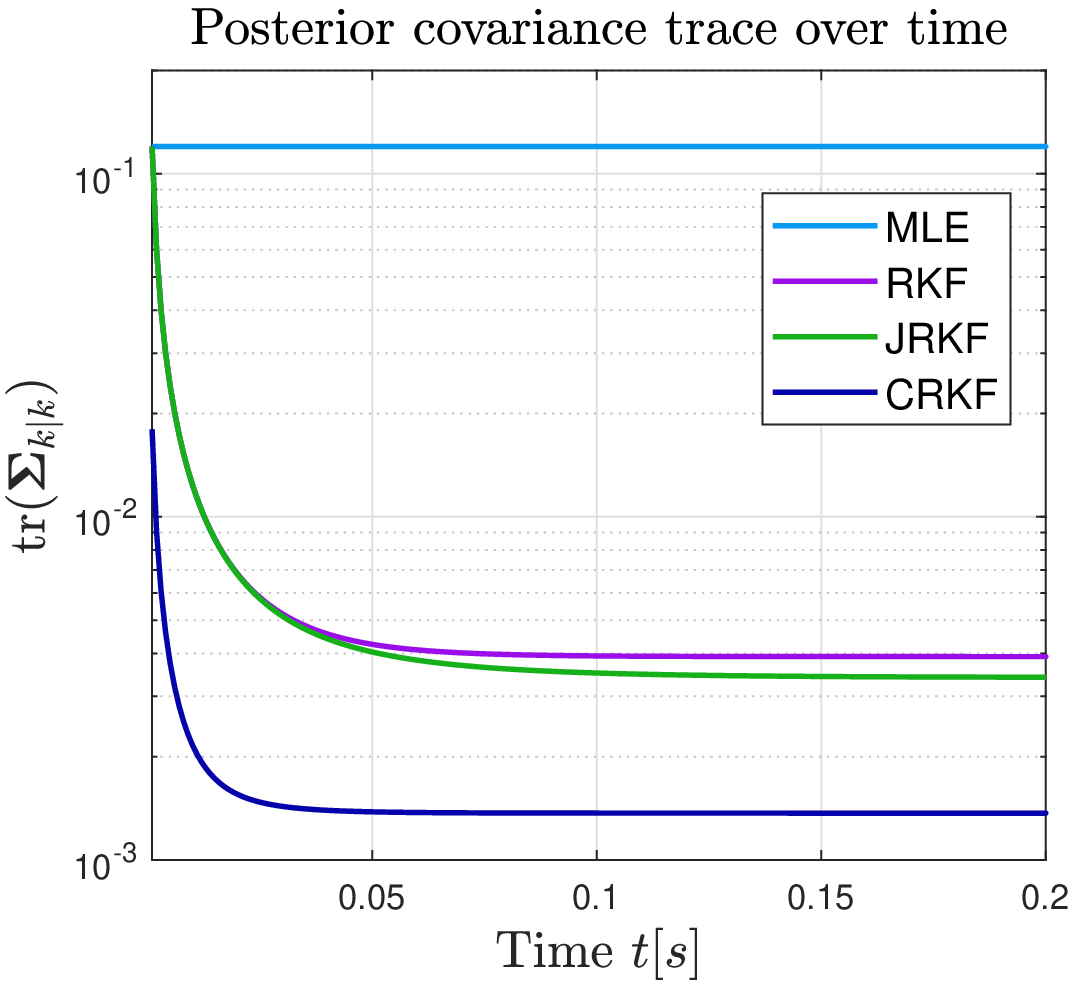}
        \caption{}
        \label{fig:tr_post_cov}
    \end{subfigure} 
    \begin{subfigure}[b]{0.49\linewidth}
        \centering
        \includegraphics[width=1.1\linewidth]{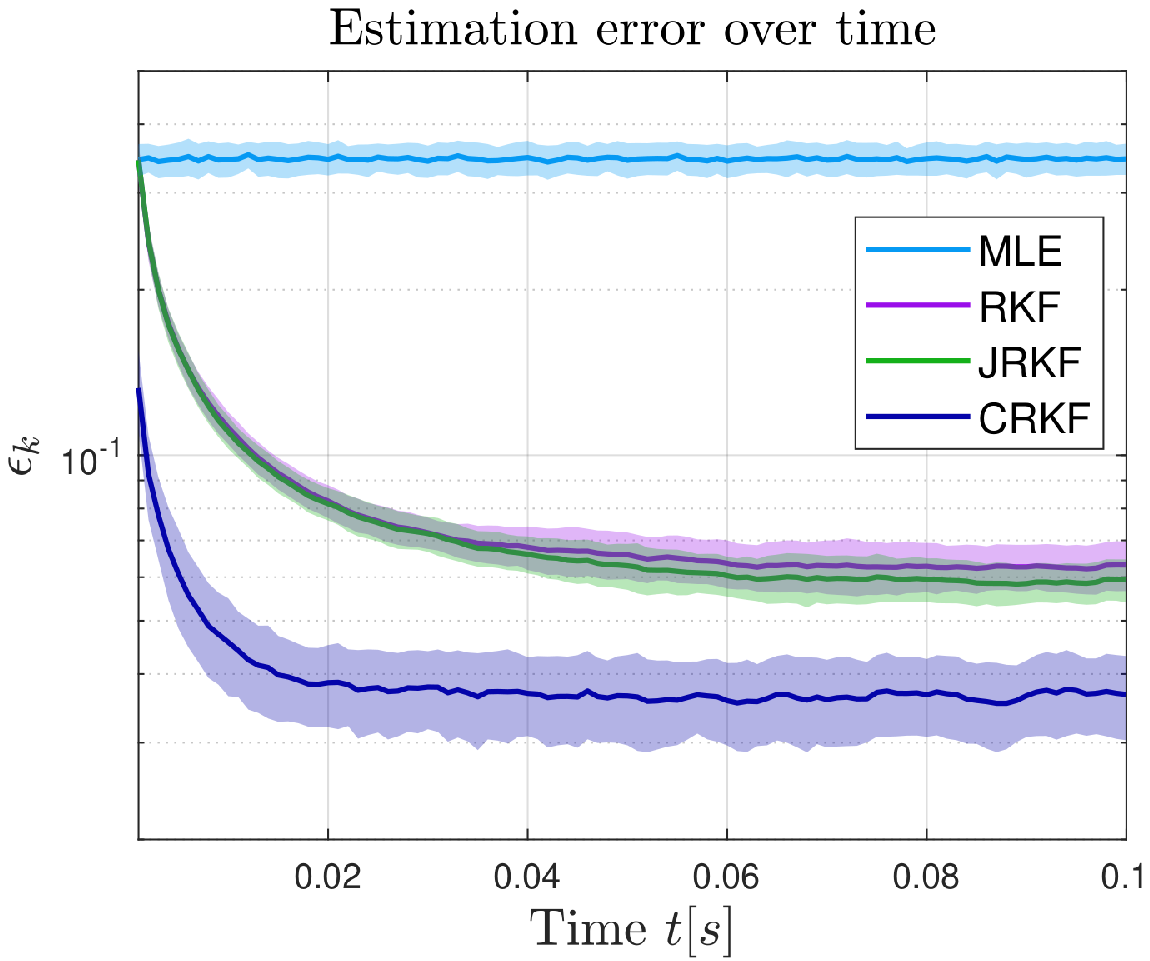}
        \caption{}
        \label{fig:tr_monte_carlo}
    \end{subfigure}
    \caption{(a) Trace of the posterior covariance over time ($\sigma_w=0.001,\sigma_v=0.1$) for the proposed estimators. (b) Estimation error over time ($\sigma_w=0.001,\sigma_v=0.1$) for the proposed estimators, where the mean and $\pm1 \sigma$ regions of the estimation errors are plotted.}
\end{figure}

To illustrate the performance of the proposed algorithms, the paths of individual agents and their final positions for a single run of the simulation are shown in \figurename\ \ref{fig:paths}. We use \cite{Lin2016} as the benchmark for the performance of the proposed algorithms. The proposed MLE and Kalman filters converge to similar formations, however the traversed paths of agents of the proposed Kalman filters are smoother compared to prevalent methods. Figure \ref{fig:tr_post_cov} shows the evolution of the posterior covariance trace over time, while the agents converge to rigid formations from their initial positions. Here, the uncertainty in edge state estimates remains constant over time for the MLE, whereas the Kalman filters use the past measurements effectively to improve estimates over time. The steady state posterior covariance trace is lowest for the CRKF, and JRFK shows an improvement over the RKF in steady state.

Furthermore, Monte Carlo simulations were performed, where for each run, the estimation error $\epsilon_k = \|\mathbf{x}_k-\hat{\mathbf{x}}_{k|k}\|_2$ is computed, which is the Euclidean norm of the edge estimation error at time step $k$. In \figurename\ \ref{fig:tr_monte_carlo}, the mean estimation error over 50 runs is shown, along with the $\pm1 \sigma$ regions. The CRKF shows the best performance in terms of estimation error, which is expected since it is the optimal filter for the global system.  Among the two proposed distributed filters, the JRKF shows significant improvement over the single-edge RKF in steady-state estimation error. Observe that in the absence of external disturbances, the edge state-space models decouple and the joint filter performs equivalently to the single-edge filter.

%% file: sections/5_summary.tex
\section{Summary}\label{sec:summary}
In this paper, we proposed statistically robust framework for estimating relative positions of agents in a formation control system. In addition to the closed form MLE, and an optimal centralized Kalman filter (CRKF), we proposed two distributed Kalman filters for both local and global state-space models. The proposed solutions can be naturally extended for any linear state-space model \cite{Williams2007}, provided the observations are a function of the relative states of agents. Moreover, to let the disturbance model better reflect real-world systems, spatio-temporal disturbance correlation may be introduced to extend the model, which naturally leads to kriged Kalman filters (see e.g. \cite{Roy2018}).

\par

